\documentclass{PoS}

\newcommand{\gsim}{\raisebox{-4pt}{$\,\stackrel{\textstyle >}{\sim}\,$}}

\title{Dilepton Production in Heavy-Ion Collisions}

\ShortTitle{Dileptons}

\author{\speaker{Ralf Rapp}
\\
        Department of Phyiscs \& Astronomy and Cyclotron Institute,
        Texas A{\&}M University, College Station, TX 77843\\
        E-mail: \email{rapp@comp.tamu.edu}}

\abstract{The properties of electromagnetic radiation from hot fireballs as 
created in ultra-relativistic heavy-ion collisions are reviewed. We first 
outline how the medium effects in the electromagnetic spectral function, 
which governs thermal production rates, relate to the (partial) restoration 
of chiral symmetry. In particular, we show how chiral and QCD sum rules, 
together with constraints from lattice QCD, can render these relations 
quantitative. Turning to dilepton data, we elaborate on updates in the 
space-time evolution and quark-gluon plasma emission rates from  
lattice-QCD calculations. With a now available excitation function in 
dilepton spectra from the RHIC beam-energy scan connecting down to 
SPS energies, we argue that a consistent interpretation of dilepton data 
emerges. Combining well-constrained space-time evolutions with state-of-the-art
emission rates identifies most of the radiation to emanate from around the 
pseudo-critical temperature, and thus confirms resonance melting as the 
prevalent mechanism in this regime, compatible with chiral restoration. 
Recent measurements of a relatively soft slope and large elliptic flow in 
direct-photon spectra at RHIC and LHC lend further support to this picture.}

\FullConference{8th International Workshop on Critical Point and Onset of Deconfinement,\\
		March 11 to 15, 2013\\
		Napa, California, USA}

\begin{document}

\section{Introduction}
Electromagnetic (EM) radiation off the expanding medium created in energetic
collision of heavy nuclei may provide a pristine glimpse at the hot 
QCD matter formed in these reactions. However, the measured spectra 
constitute, radiation yields integrated over the entire lifetime 
of the fireball. To effectively discriminate the different components in the 
spectra (e.g., primordial production, early QGP radiation, hot/dense matter 
around $T_c$ and late hadronic emission), the full richness of this 
observable needs to be exploited. In this regard, and advantageous feature of 
the local thermal  emission rate, 
\begin{equation}
\frac{dN_{ll}}{d^4xd^4q} = \frac{\alpha_{\rm EM}^2 L(M)}{6\pi^3 M^2} \
f^B(q_0;T) \ \rho_{\rm EM}(M,q;\mu_B,T) \ ,
\label{rate}
\end{equation}
is its separate dependence on the invariant mass ($M$) and 3-momentum ($q$) 
through the vector spectral function, $\rho_{\rm EM}$, of the medium. On the 
one hand, the $M$-dependence encodes the dynamical effects of the microscopic 
interactions governing its shape (e.g., information on the degrees of freedom 
or chiral symmetry restoration). On the other hand, the momentum dependence 
mostly probes the kinematics of the medium, i.e., the interplay of decreasing 
temperature and increasing blue shift in the fireball expansion, well-known 
from hadron spectra at thermal freeze-out. For EM radiation, this interplay 
is encoded in a superposition of all phases of the fireball, and thus requires 
further disentangling. The main lever arm here is the competition between 
the Bose factor, $f^B$, favoring early phases, and the emitting 3-volume, 
$V_{\rm FB}$, favoring late phases. Since the temperature sensitivity of the 
Bose factor increases exponentially at large energies (i.e., large mass, 
large 3-momentum, or both), the latter prevail at high temperatures, while 
the weaker sensitivity at small $q_0$ shifts the main emission to smaller 
temperatures where the radiating volume grows with an inverse power in $T$. 
At a quantitative level, the temperature/density dependence of the spectral 
function also figures into these considerations, especially in the low-mass 
regime (LMR, $M\le 1$\,GeV). Another valuable diagnostic tool that has recently 
become available in the EM sector is the elliptic flow. Since for EM spectra
the $v_2$ is a weighted sum over all phases, its magnitude, relative
to the final-state hadrons, can serve as another indicator of the emission
time; in typical hydrodynamic evolutions at RHIC, the bulk elliptic flow
takes about 5\,fm/$c$ to develop most of its strength. Clearly, robust 
interpretations of the spectral shape of the emitted radiation need to be 
in concert with a good control over the emission temperatures.

In the following, we will first give an update on implications of
hadronic medium effects in the vector spectral function for chiral 
restoration and introduce Quark-Gluon Plasma (QGP) 
emission rates motivated by thermal lattice-QCD (lQCD) computations 
(Sec.~\ref{sec_rates}). This also raises the issue of consistency of 
microscopic emission rates and the equation of state (EoS) governing the 
bulk evolution, in particular the local temperature.
We will then turn to a discussion of low-mass dilepton spectra from SPS
to top RHIC energy, and address recent measurements of spectra and 
elliptic flow of direct photons at RHIC and LHC (Sec.~\ref{sec_spectra}).  
A brief conclusion is given in Sec.~\ref{sec_concl}.

\section{Thermal EM Emission Rates and Chiral Symmetry Restoration}
\label{sec_rates}
\begin{figure}[!tbp]
\begin{minipage}{13pc}
\includegraphics[width=1.0\textwidth]{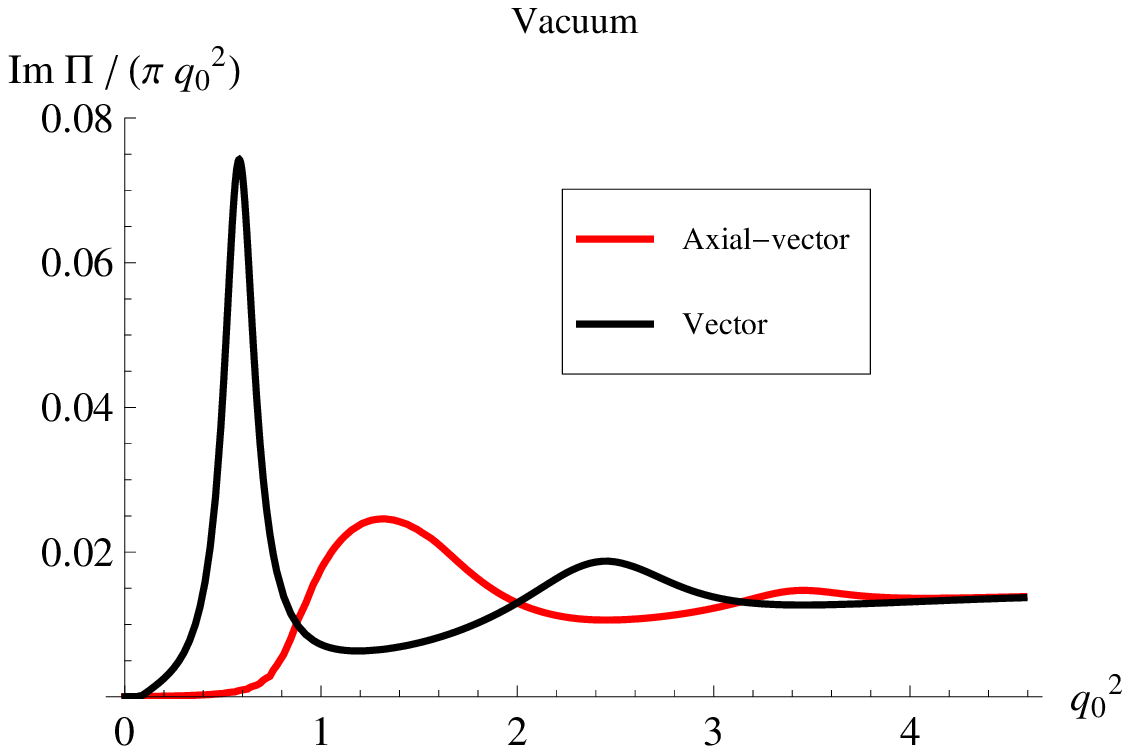}
\end{minipage}
\hspace{-2.5pc}
\begin{minipage}{13pc}
\includegraphics[width=1.0\textwidth]{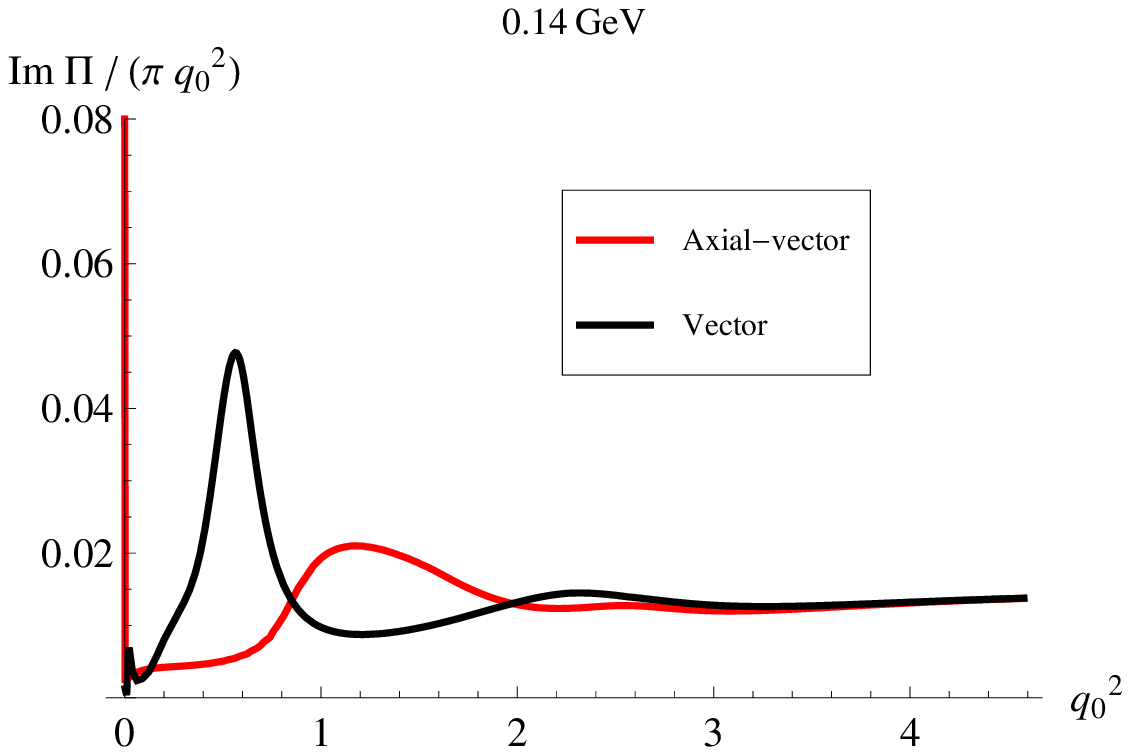}
\end{minipage}
\hspace{-2.5pc}
\begin{minipage}{13pc}
\includegraphics[width=1.0\textwidth]{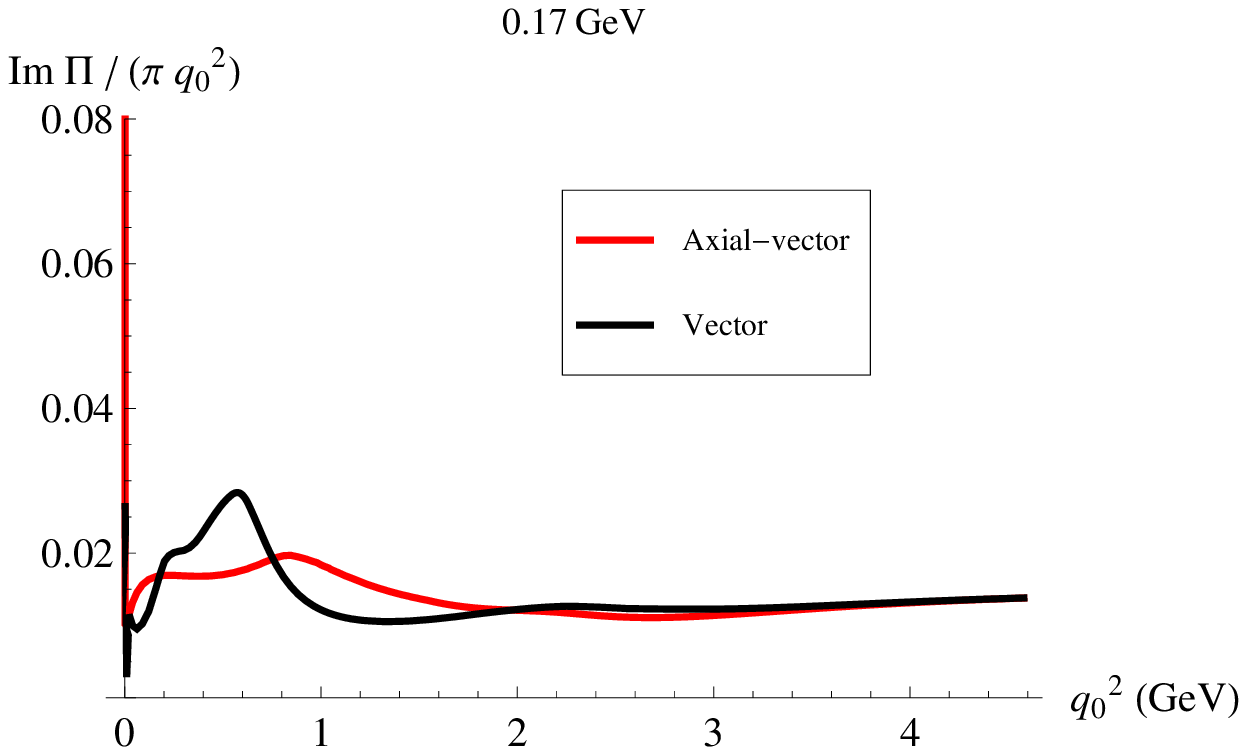}
\end{minipage}\hspace{0pc}
\caption{Isovector-vector and -axialvector spectral functions in vacuum (left 
panel) and at finite temperature (middle panel: $T$=140\,MeV, right panel: 
$T$=170\,MeV)~\cite{Hohler:2012fj}.}
\label{fig_VAspec}
\end{figure}

\begin{figure}[!tbp]
\vspace{-0.5cm}
\begin{minipage}{18pc}
\includegraphics[width=0.92\textwidth]{drdm2-Inx160.eps}
\end{minipage}
\hspace{0.1pc}%
\begin{minipage}{18pc}
\includegraphics[width=0.87\textwidth,angle=-90]{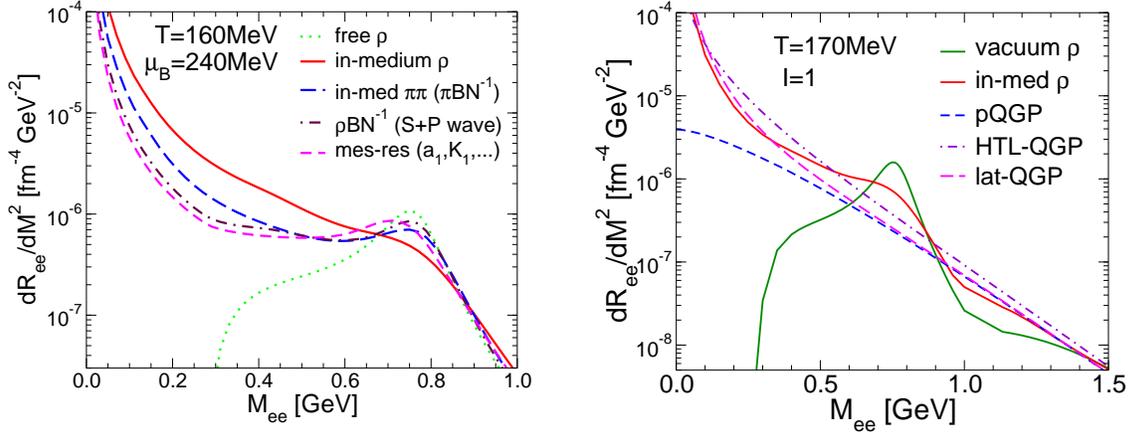}
\end{minipage}
\vspace{-0.5cm}
\caption{Thermal dilepton rates (integrated over 3-momentum) in hot/dense 
matter characteristic for top 
SPS energy (left panel) and top RHIC energy (right panel). The former shows
the total hadronic rate and its decomposition into pion-cloud and resonance
channels while the latter compares the total vacuum and in-medium hadronic 
rates to QGP emission within the hard-thermal-loop 
approach~\cite{Braaten:1990wp} and lQCD~\cite{Ding:2010ga}.}
\label{fig_rates}
\end{figure}

Effective hadronic Lagrangians, combining chiral pion interactions with 
resonance excitations, implemented into finite-temperature field theory,
have been widely applied to evaluate vector-meson spectral functions 
in hot and/or dense hadronic matter, see, e.g., 
Ref.~\cite{Rapp:2009yu,Leupold:2009kz} for recent reviews.   
The generic outcome of these calculations is an appreciable broadening
of the spectral peaks with little, if any, mass shift, provided that the
parameters of the vacuum Lagrangian (coupling constants and bare masses)
are not subject to in-medium changes.
For the $\rho$ meson, the broadening amounts to a few hundred MeV at hadronic
densities of $\varrho_h$=0.2\,fm$^{-3}$, 
leading to its melting when extrapolated into
the regime of the expected QCD phase boundary ($T_{\rm pc}$$\simeq$\,170MeV),
cf.~the black lines in Fig.~\ref{fig_VAspec}. The dissolution of the
hadronic resonance structure suggests a change of the relevant degrees of
freedom in the system, and thus may be interpreted as an indicator of 
deconfinement~\cite{Dominguez:1989bz}.
Another issue is if and how these medium effects signal the restoration of the
spontaneously broken chiral symmetry. This is quantified by Weinberg sum 
rules (WSRs)~\cite{Weinberg:1967,Das:1967ek}, 
\begin{equation}
  f_n = - \int\limits_0^\infty ds \ s^n \
  \left[\rho_V(s) - \rho_A(s) \right]  \ ,
\label{wsr}
\end{equation}
which relate moments of the difference between the isovector-vector
and -axialvector spectral function to order parameters 
of chiral breaking, 
$f_{-2} = f_\pi^2 \langle r_\pi^2\rangle/3 - F_A$, $f_{-1} = f_\pi^2$, 
$f_0   = f_\pi^2 m_\pi^2$, $f_1 = -2\pi \alpha_s \langle {\cal O}_4^\chi \rangle$ 
($r_\pi$: pion charge radius, $F_A$: coupling constant for the radiative
pion decay, $\langle {\cal O}_4^\chi \rangle$: chirally breaking 4-quark 
condensate). These sum rules remain valid at finite 
temperature~\cite{Kapusta:1993hq}, independently at each 3-momentum for 
longitudinal and transverse components. Ideally, one would compute both 
in-medium spectral function in 
a chiral approach, evaluate the integrals and test for agreement with the 
order parameters, as given, e.g., by thermal lQCD. However, 
realistic calculations of the in-medium axialvector spectral function are
not yet available. In Ref.~\cite{Hohler:2012fj}, a more modest question has 
been addressed, namely whether solutions to Weinberg and QCD sum rules, with 
order parameters from lQCD as available, can be found using existing 
in-medium calculations of the $\rho$ spectral function~\cite{Hohler:2012xd}.
Toward this end, a quantitative fit to the axial-/vector $\tau$ data
was employed which accurately satisfies the sum rules in vacuum. In this 
fit, Breit-Wigner ans\"atze for the $a_1$ and excited resonances
($\rho'$ and $a_1'$) have been used, and their masses and widths were 
required to change monotonically with temperature. 
A viable solution was indeed found, with the resulting spectral functions
clearly exhibiting their mutual approach toward degeneracy, see 
Fig.~\ref{fig_VAspec}. While this is not a proof of chiral restoration, it
nevertheless shows that a strongly broadened $\rho$ spectral function,
as will be used in applications to dilepton data below, is {\em compatible} 
with it. Another indication for this compatibility arises from the realization 
that the processes generating the $\rho$ 
broadening (resonances and pion cloud modifications) find their 
counterparts in reducing the chiral condensate. In dilute matter, the latter 
decreases according to~\cite{Gerber:1988tt}
\begin{equation}
\frac{\langle\bar qq\rangle (T,\mu_B)}{\langle \bar qq\rangle_0} \ = \
1-\sum\limits_h \frac{\varrho_h^s \Sigma_h}{m_\pi^2 f_\pi^2}
\label{qqbar-med}
\end{equation}
($\varrho_h^s$: scalar density), where 
$\Sigma_h = m_q \langle h|\bar qq|h\rangle$ is characterized by the scalar 
quark number inside hadron $h$; it can be decomposed into contributions from 
its quark core and pion cloud ~\cite{Jameson:1992,Birse:1992}, 
$\Sigma_h = \Sigma_h^{\rm core} + \Sigma_h^{\pi}$, which are roughly equal 
in magnitude (in analogy to the medium effects in the dilepton rate, 
see left panel of Fig.~\ref{fig_rates}). The resulting suppression of the 
quark condensate from a (non-interacting) hadron resonance gas reproduces 
lQCD calculations rather well~\cite{Borsanyi:2010bp}. 

Progress has been made in extracting the QGP emission rate from 
lQCD at zero pair momentum, $q$=0~\cite{Ding:2010ga,Brandt:2012jc}. 
Applications to experiment require 
the extension to finite $q$; in Ref.~\cite{Rapp:2013nxa} this has been 
constructed by implementing the $q$=$q_0$ dependence of the perturbative
photon rate and matching it to the conductivity in lQCD. The  
$q$-integrated rates are similar in shape to the hard-thermal-loop results,
albeit quantitatively somewhat smaller in the LMR. One reason for this could 
be the smaller number of degrees of freedom that characterize the 
(nonperturbative) QGP in lQCD relative to the perturbative system 
underlying the HTL rates. It is thus important to maintain consistency 
between emission rates and EoS in applications to 
heavy-ion collisions (as will be done below).
One also sees from the right panel of Fig.~\ref{fig_rates} that, for temperatures
around $T$=170\,MeV, the bottom-up extrapolated in-medium hadronic rates 
approximately coincide with the top-down extrapolated QGP rates.     

Microscopic calculations of in-medium dilepton rates can be straightforwardly
carried to the photon point, i.e., $M$=0 and $q_0$=$q$. For hadronic rates
this has been done in Ref.~\cite{Turbide:2003si}, where, in addition, $\omega$
$t$-channel exchange in $\pi\rho\to\pi\gamma$ has been identified as an important
photon source at $q_t\gsim 1.5$\,GeV. It turns out that the hadronic emission
rate is quite comparable to the complete leading-order QGP
rate~\cite{Arnold:2001ms} in the vicinity of $T_{\rm pc}$.

\section{EM Spectra in Heavy-Ion Collisions}
\label{sec_spectra}
In the following we confront the models of in-medium hadronic and QGP rates 
(as discussed above), after folding over thermal fireballs constrained by 
hadron data (yields, spectra and $v_2$), to dilepton (Sec.~\ref{sec_dilep}) and
photon data (Sec.~\ref{sec_phot}) at SPS, RHIC and LHC.

\subsection{Low-Mass Dileptons}
\label{sec_dilep}
\begin{figure}[!tbp]
\vspace{-0.3cm}
\begin{minipage}{18pc}
\vspace{-0.4cm}
\includegraphics[width=0.98\textwidth]{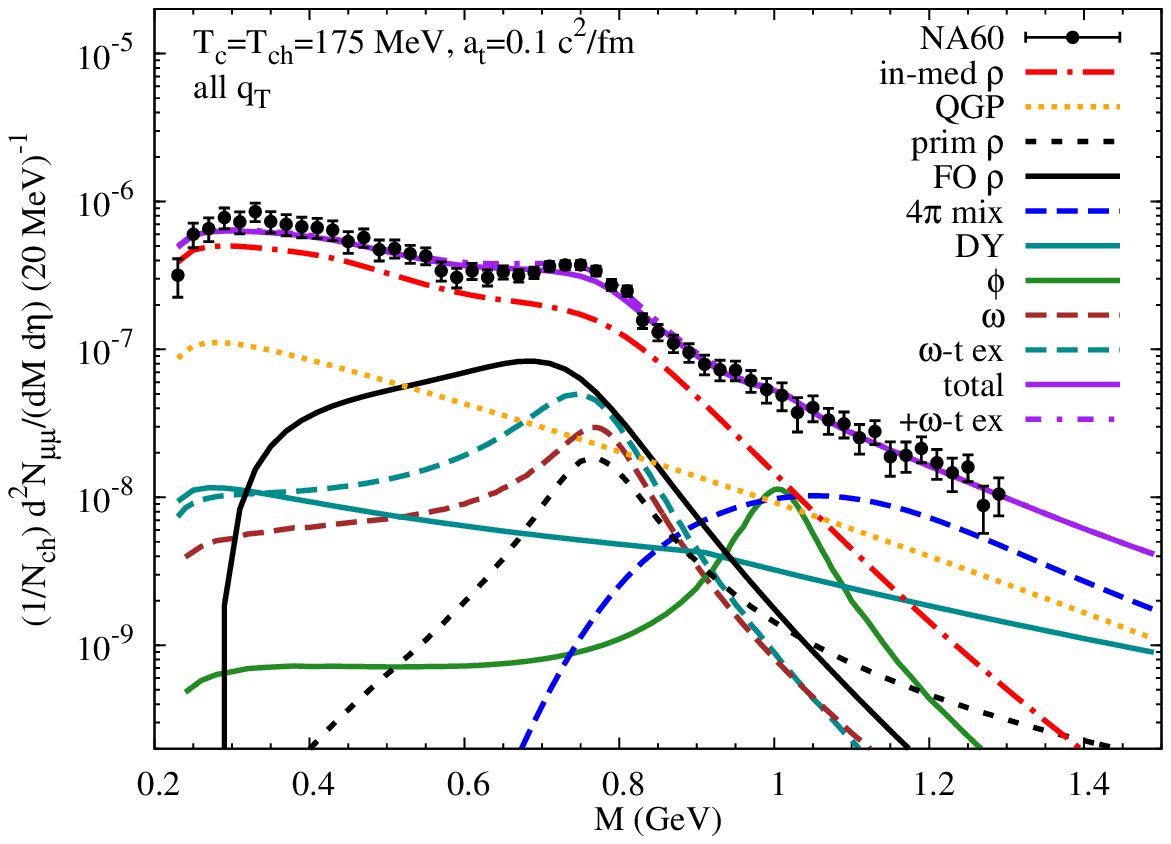}
\end{minipage}
\hspace{0.5pc}%
\begin{minipage}{18pc}
\vspace{-0.2cm}
\includegraphics[width=0.86\textwidth,height=1.15\textwidth,angle=-90]{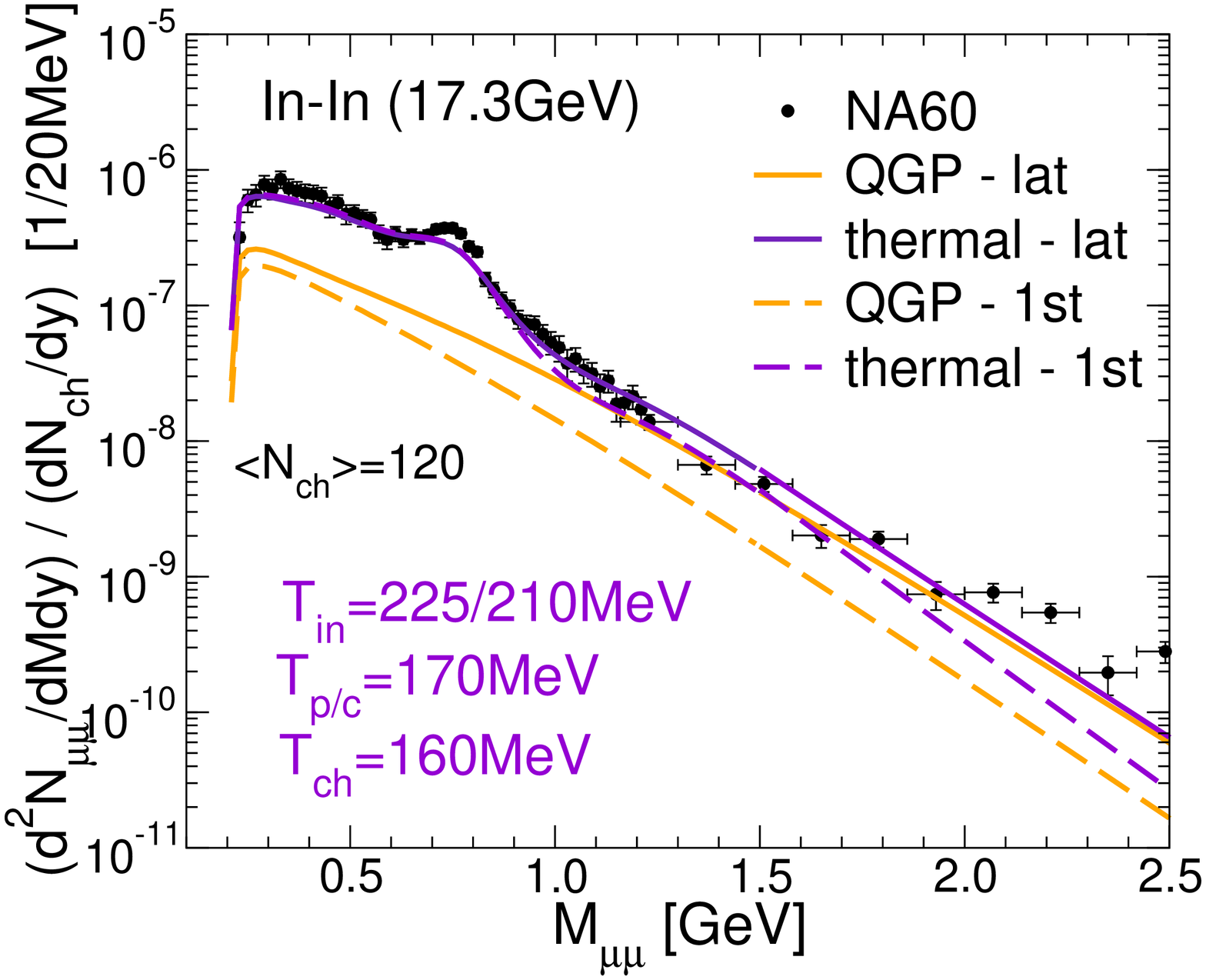}
\end{minipage}
\vspace{-0.4cm}
\caption{Dimuon excess spectra in In-In($\sqrt{s}$=17.3\,GeV) collisions at SPS 
as measured by NA60 (acceptance corrected)~\cite{Arnaldi:2006jq}, compared to 
theoretical calculations with in-medium vector spectral functions (left 
panel)~\cite{Rapp:2009yu}, and illustrating their sensitivity to the QGP EoS
(right panel).}
\label{fig_na60}
\end{figure}
In the NA60 dimuon excess spectra~\cite{Arnaldi:2006jq} the 
contributions from final-state hadron
decays could be subtracted thanks to excellent statistics and mass 
resolution. In the LMR, the predictions of a melting $\rho$ resonance
agree well with the data, while continuum radiation from multi-hadron 
annihilation and the QGP figures for $M$>1\,GeV (Fig.~\ref{fig_na60} left). 
The relative contributions and spectral shape (as given by the in-medium 
spectral function and overall Bose factor) in the {\it invariant}-mass spectra 
are determined by the temperature profile of the fireball, cooling 
from $T_i$$\simeq$200\,MeV  to $T_{\rm fo}$$\simeq$120\,MeV, with little 
sensitivity to the expansion dynamics. The overall yield could not be 
accurately predicted, but rather allows to
extract the (average) fireball lifetime as $\tau_{\rm FB}$=6.5$\pm$1\,fm/$c$.
The impact of replacing a first-order by a lQCD EoS is illustrated 
in Fig.~\ref{fig_na60} right. In the LMR, the QGP yield increases at the
expense of the hadronic one, resulting in a very similar total. However, at 
intermediate mass, larger QGP temperatures 
resulting from the nonperturbative reduction of the lattice EoS increase 
the total yield significantly, which is favored by the data.
 
\begin{figure}[!tbp]
\begin{minipage}{18pc}
\includegraphics[width=0.95\textwidth]{dndm-ceres40.eps}
\end{minipage}
\hspace{0.5pc}%
\begin{minipage}{18pc}
\vspace{-0.2cm}
\includegraphics[width=0.95\textwidth]{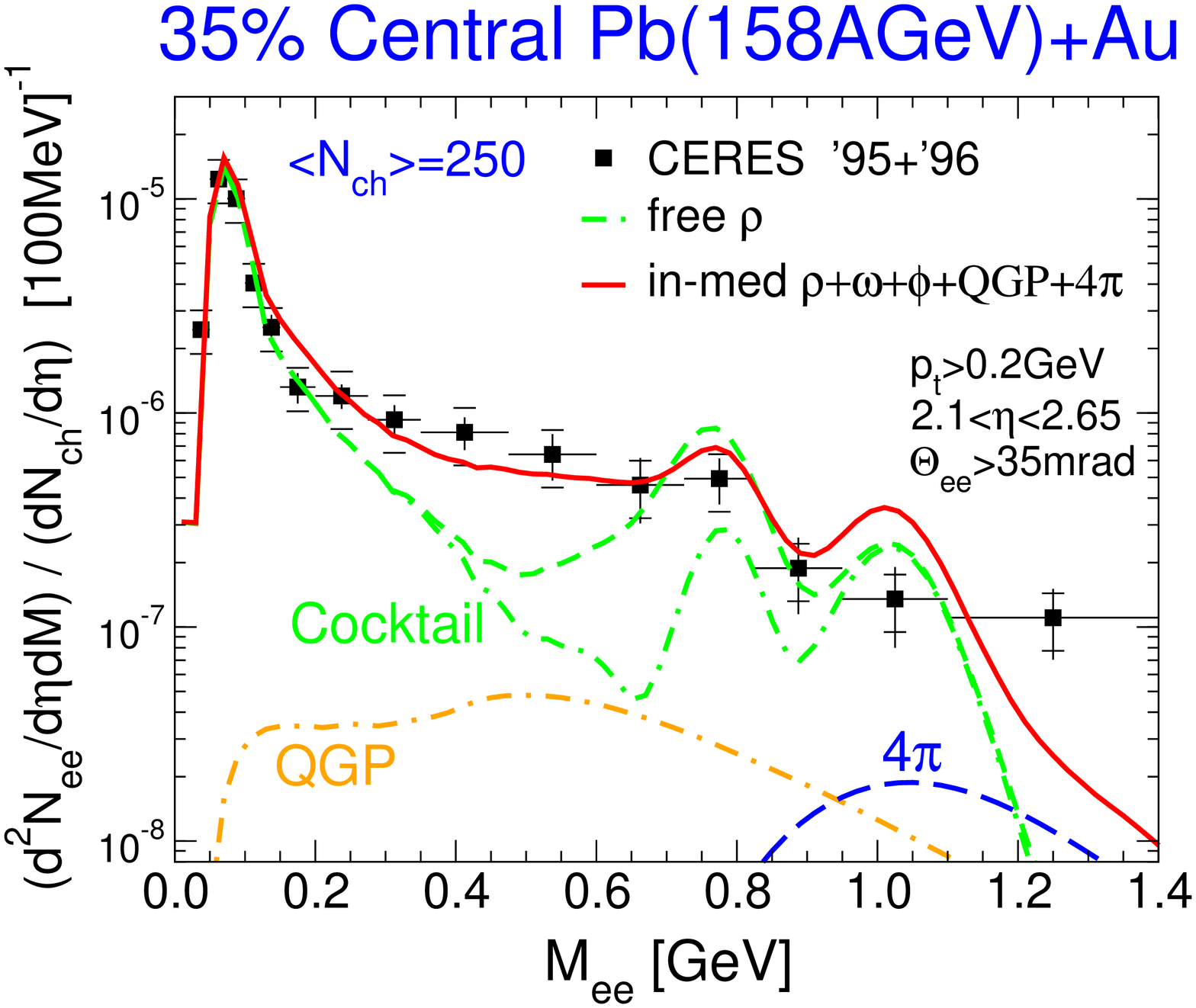}
\end{minipage}
\begin{center}
\begin{minipage}{33pc}
\includegraphics[width=0.98\textwidth]{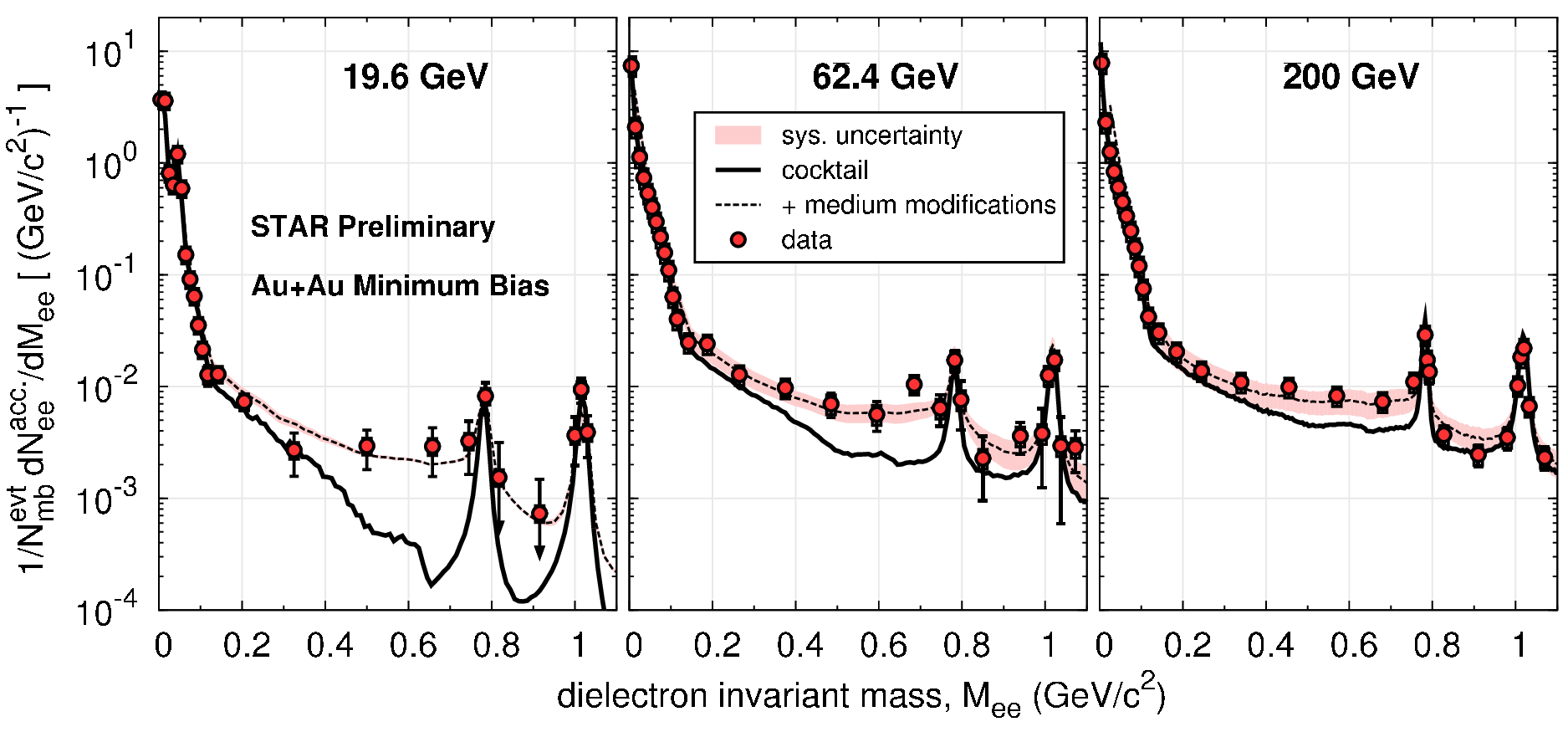}
\end{minipage}
\end{center}
\caption{Excitation function of dielectron spectra from 
CERES/NA45~\cite{Agakichiev:2005ai} at 
$\sqrt{s}$=8.8\,GeV (upper left) and 17.3\,GeV (upper right) and from 
STAR at $\sqrt{s}$=19.6, 62.4 and 200\,GeV (lower panels)~\cite{Geurts:2012rv}.}
\label{fig_excit}
\end{figure}
Dielectron measurements from the SPS and RHIC are summarized in 
Fig.~\ref{fig_excit}. The recent STAR data from the RHIC beam-energy 
scan~\cite{Geurts:2012rv} constitute a first systematic excitation function, 
establishing consistency with previous SPS results~\cite{Agakichiev:2005ai}. 
The strongly broadened $\rho$ spectral function plus a moderate QGP 
contribution, as found at the SPS, describe the data up to top RHIC energy. 
This indicates a universal emission source, with large contributions from 
around $T_{\rm pc}$ and 
hadronic medium effects driven by baryons and antibaryons. 
We recall, however, that these calculations cannot explain the large low-mass, 
low-momentum enhancement observed by PHENIX in central Au-Au, while the central
STAR data tend to be slightly overestimated around $M$$\simeq$0.2\,GeV. 

\subsection{Direct Photons}
\label{sec_phot}
Direct photon radiation has been measured at RHIC and LHC and also shows a
substantial excess over primordial and final-state hadron decay sources. The
excess spectra carry inverse slopes of 
$T_{\rm eff}=221\pm27$\,MeV (RHIC)~\cite{Adare:2008ab} and $301\pm51$\,MeV 
(LHC)~\cite{Wilde:2012wc}, and an appreciable $v_2$. The latter is difficult
to explain by early QGP radiation, but the inverse slopes actually point
at ``later" emission as well. Using the blue-shift expression, 
$T_{\rm eff}\simeq T \sqrt{(1+\beta)/(1-\beta)}$, with an average radial
flow velocity of $\beta$=0.3-0.4, leads to emission temperatures of
$T$$\simeq$160-200\,MeV and renders a large $v_2$ plausible. 
Nevertheless, it is not easily reproduced in realistic 
calculations.   

\begin{figure}[!tbp]
\begin{minipage}{18pc}
\vspace{-0.4cm}
\includegraphics[width=0.98\textwidth]{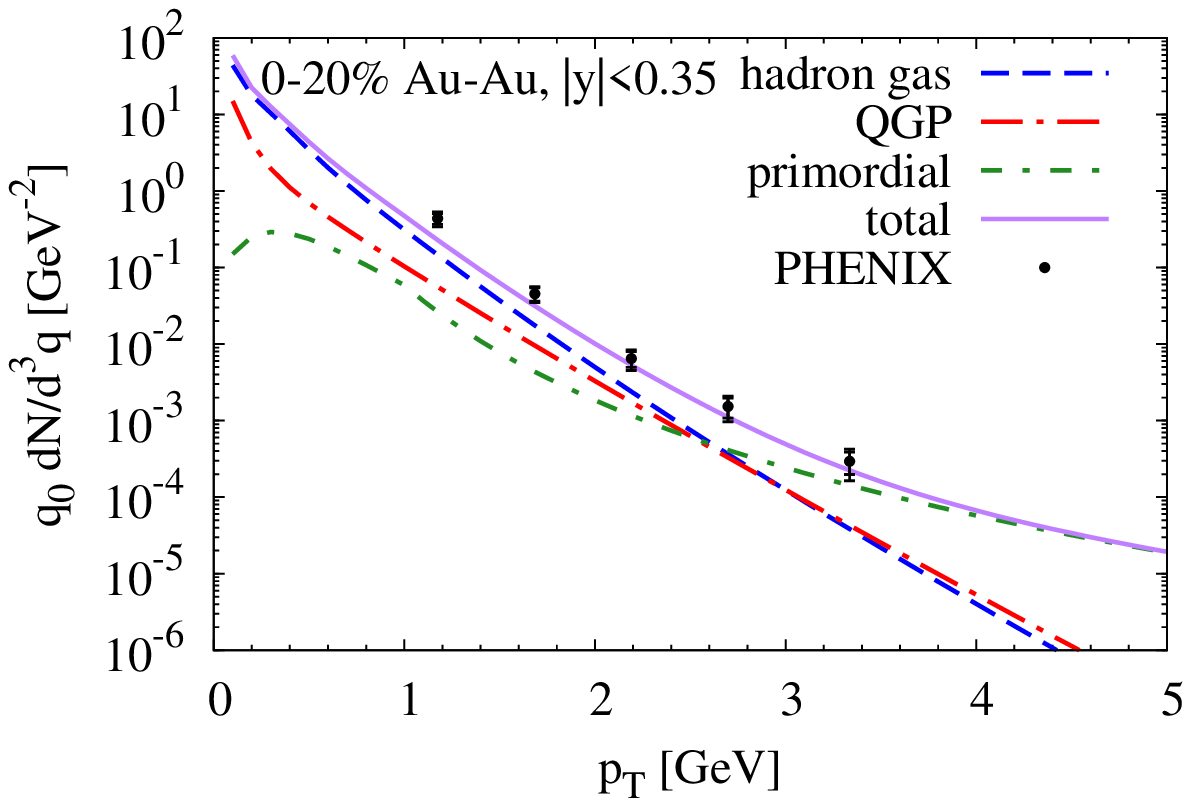}
\end{minipage}
\hspace{0.5pc}%
\begin{minipage}{18pc}
\vspace{-0.2cm}
\includegraphics[width=0.98\textwidth]{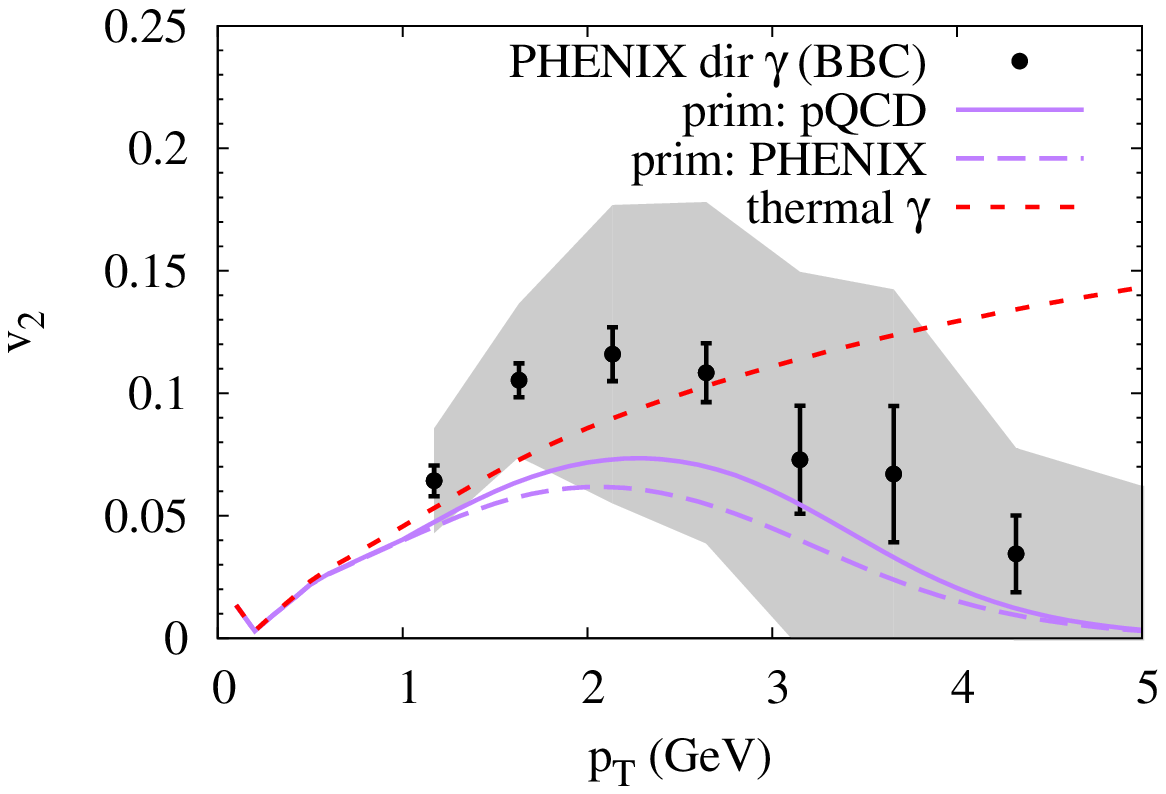}
\end{minipage}\hspace{0pc}
\vspace{-0.0cm}
\begin{minipage}{18pc}
\vspace{-0.0cm}
\includegraphics[width=0.73\textwidth,height=1.04\textwidth,angle=-90]{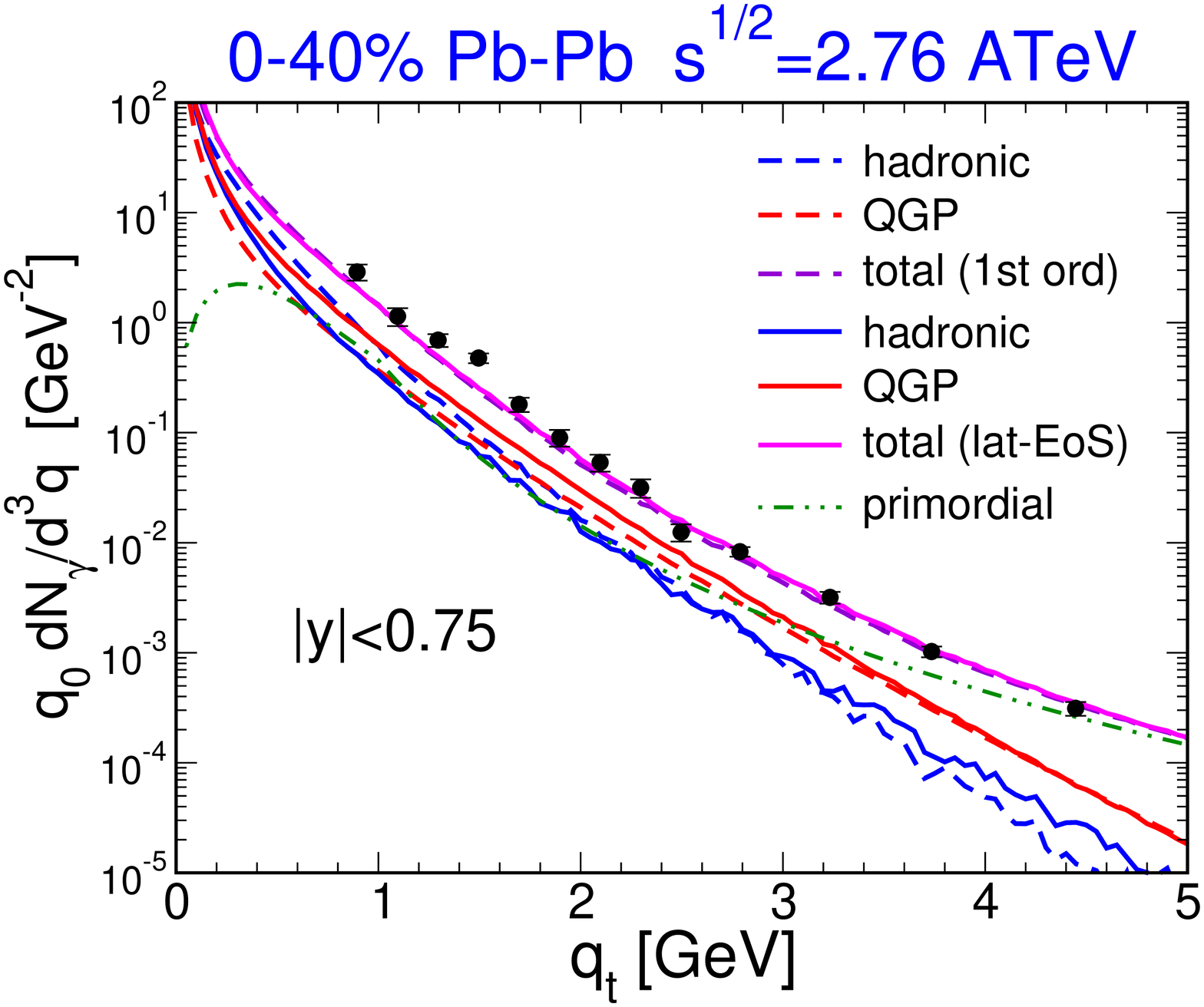}
\end{minipage}
\hspace{0.5pc}%
\begin{minipage}{18pc}
\vspace{0.4cm}
\includegraphics[width=0.98\textwidth]{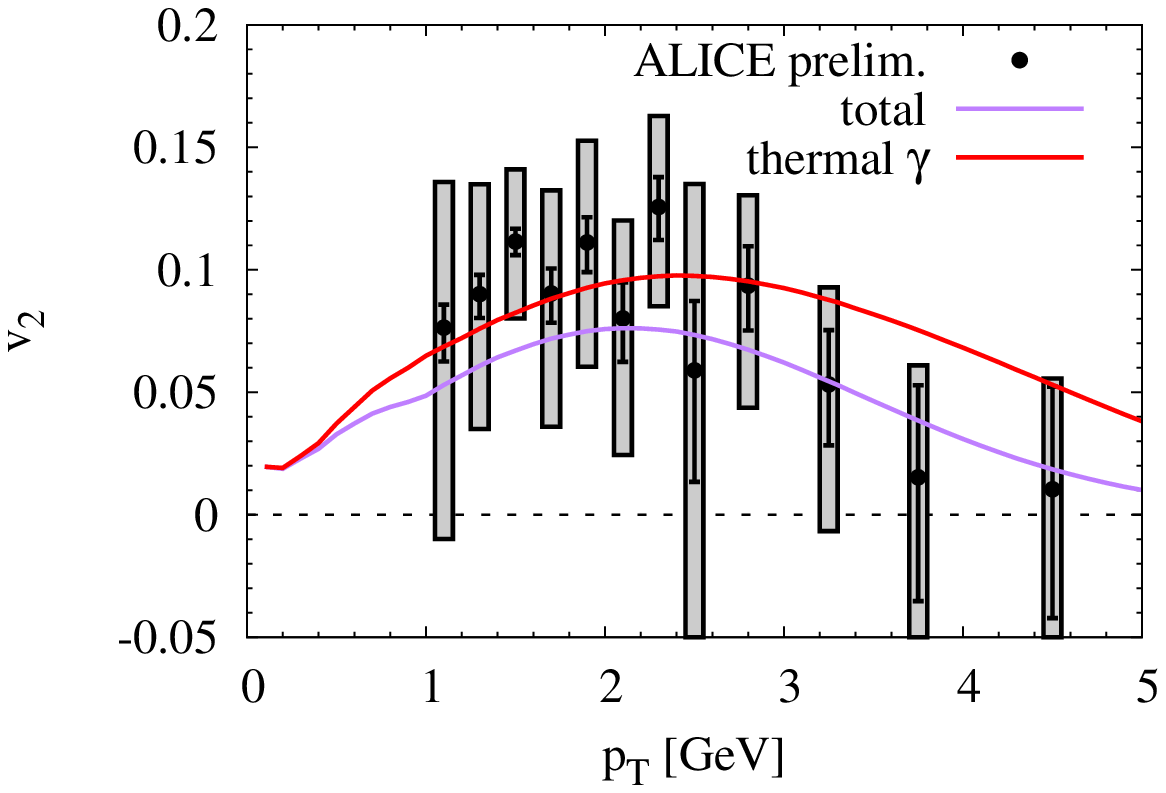}
\end{minipage}\hspace{0pc}
\caption{Direct photon spectra (left panels) and elliptic flow (right panels) in 0-20\% 
Au-Au($\sqrt{s}$=200\,GeV) at RHIC (upper panels) and 0-40\% PbPb($\sqrt{s}$=2.76\,TeV) at 
LHC (lower panels).}
\label{fig_gam}
\end{figure}

In Ref.~\cite{vanHees:2011vb}, LO QGP and in-medium hadronic rates have been
evolved over a thermal fireball model. The latter has been constrained by
measured hadron spectra and $v_2$, with a key feature of the bulk-$v_2$
leveling off around $T_{\rm pc}$, i.e., after ca.~5-6~fm/$c$ into the evolution
of Au-Au at RHIC. This is not necessarily the case in hydrodynamic
simulations~\cite{Chatterjee:2009ys,Dion:2011pp}, but can be realized when 
utilizing a non-zero initial flow
together with a rather compact initial energy density profile~\cite{He:2011zx}.
This, in particular, leads to the realization of ``sequential freezeout", i.e., 
the kinetic decoupling of multi-strange hadrons ($\phi$, $\Xi$, $\Omega^-$)
at chemical freezeout, $T_{\rm ch} \simeq T_{\rm pc}$. The resulting thermal
photon spectra lead to approximate agreement with the PHENIX data, while
the $v_2$ is at the lower end of the experimental uncertainty. At the LHC,
the agreement with preliminary data in 0-40\% Pb-Pb(2.76\,ATeV) is fair.
The $q_t$ spectra illustrate a significant reshuffling of QGP and hadronic
contributions when switching from first-order to lQCD EoS, reiterating
the large contributions from around $T_{\rm pc}$.

\section{Conclusions}
\label{sec_concl}
Electromagnetic radiation in heavy-ion collisions continues to illuminate the 
properties of the formed medium. Low-mass dilepton spectra and their 
interpretation are developing into 
a rather consistent picture, where the melting of the $\rho$ meson established 
at SPS seems to prevail also at RHIC. We have argued that this melting is 
theoretically compatible with chiral symmetry restoration and suggestive for a 
gradual change in the effective degrees of freedom in the system. Taken together 
with the temperature slopes extracted from the invariant-mass spectra, we may 
well have evidence for the long-sought for radiation from the QCD transition 
region. The inverse slopes and remarkable $v_2$ in the direct photons support 
this interpretation, even though a full theoretical understanding has not yet 
been achieved (possibly calling for additionally enhanced photon rates around 
$T_{\rm pc}$, and/or initial-state effects~\cite{Bzdak:2012fr}). 
Clearly, a dilepton $v_2$ measurement, as well as precision 
mass spectra at collider energies, are needed to further test and deepen 
our understanding.


\noindent {\bf Acknowledgment}\\
This work has been supported by the U.S. National Science Foundation
under grants no.~PHY-0969394 and PHY-1306359, and by the A.-v.-Humboldt 
Foundation.


\begin{thebibliography}{99}
\bibitem{Rapp:2009yu}
  R.~Rapp, J.~Wambach and H.~van Hees,
  in {\em Relativistic Heavy-Ion Physics}, edited by R.~Stock and
  Landolt B\"ornstein (Springer), New Series {\bf I/23A} (2010) 4-1
  [arXiv:0901.3289[hep-ph]].

\bibitem{Leupold:2009kz}
  S.~Leupold, V.~Metag and U.~Mosel,
  Int.\ J.\ Mod.\ Phys.\  E {\bf 19}, 147 (2010).

\bibitem{Hohler:2012fj}
  P.M.~Hohler and R.~Rapp,
  EPJ Web Conf. {\bf 36}, 00012 (2012); in preparation (2013).

\bibitem{Dominguez:1989bz}
  C.A.~Dominguez and M.~Loewe,
  Phys.\ Lett.\ B {\bf 233}, 201 (1989).

\bibitem{Weinberg:1967}
S.~Weinberg, Phys. Rev. Lett. \textbf{18}, 507 (1967).
                                                                                
\bibitem{Das:1967ek}
T.~Das, V.S.~Mathur, and S.~Okubo, Phys. Rev. Lett. \textbf{19}, 859 (1967).
                                                                                
\bibitem{Kapusta:1993hq}
  J.I.~Kapusta and E.V.~Shuryak,
  Phys.\ Rev.\ D {\bf 49}, 4694 (1994).

\bibitem{Hohler:2012xd}
  P.M.~Hohler and R.~Rapp,
  Nucl. Phys. A {\bf 892},  58 (2012).

\bibitem{Gerber:1988tt}
  P.~Gerber and H.~Leutwyler,
  Nucl.\ Phys.\ B {\bf 321}, 387 (1989).
                                                                                
                                                                                

\bibitem{Jameson:1992}
I.~Jameson, A.W.~Thomas and G.~Chanfray, J. Phys. G {\bf 18}, L159 (1992).
                                                                                
\bibitem{Birse:1992}
M.C.~Birse and J.A.~McGovern, Phys. Lett. {\bf B292}, 242 (1992).

\bibitem{Braaten:1990wp}
E.~Braaten, R.D.~Pisarski, and T.-C.~Yuan, Phys. Rev. Lett. \textbf{64}, 
2242 (1990).

\bibitem{Ding:2010ga}
  H.T.~Ding {\it et al.}, 
  Phys.\ Rev.\  D {\bf 83}, 034504 (2011).

\bibitem{Borsanyi:2010bp}
S.~Borsanyi {\it et al.}
[Wuppertal-Budapest Coll.],
  JHEP {\bf 1009}, 073 (2010).

\bibitem{Brandt:2012jc} 
  B.B.~Brandt, A.~Francis, H.B.~Meyer and H.~Wittig,
  JHEP {\bf 1303}, 100 (2013).

\bibitem{Rapp:2013nxa}
  R.~Rapp,
  arXiv:1304.2309 [hep-ph].

\bibitem{Turbide:2003si}
  S.~Turbide, R.~Rapp and C.~Gale,
  Phys.\ Rev.\  C {\bf 69}, 014903 (2004).

\bibitem{Arnold:2001ms}
P.B.~Arnold, G.D.~Moore and L.G.~Yaffe,
  JHEP {\bf 0112}, 009 (2001).

\bibitem{Arnaldi:2006jq}
R.~Arnaldi et~al. [NA60 Coll.], 
Phys. Rev. Lett. \textbf{96}, 162302 (2006); {\it Eur. Phys. J. C} {\bf 61}, 711 (2009).

\bibitem{Agakichiev:2005ai}
  G.~Agakichiev {\it et al.}  [NA45 Coll.],
  Eur. Phys. J. {\bf C41}, 475 (2005); Phys. Rev. Lett. \textbf{91},
042301 (2003).

\bibitem{Geurts:2012rv} 
  F.~Geurts {\it et al.} [STAR Coll.],
  Nucl.\ Phys.\ A904-905 {\bf 2013}, 217c (2013).

\bibitem{Adare:2008ab} 
  A.~Adare {\it et al.}  [PHENIX Coll.],
  Phys.\ Rev.\ Lett.\  {\bf 104}, 132301 (2010); 
 {\it ibid.}  {\bf 109}, 122302 (2012).

\bibitem{Wilde:2012wc} 
  M.~Wilde {\it et al.} [ALICE Coll.],
  Nucl.\ Phys.\ {\bf A904-905}, 573c (2013); arXiv:1212.3995 [hep-ex].


\bibitem{vanHees:2011vb}
  H.~van Hees, C.~Gale and R.~Rapp,
  Phys.\ Rev.\ C {\bf 84} (2011) 054906.

\bibitem{He:2011zx} 
  M.~He, R.J.~Fries and R.~Rapp,
  Phys.\ Rev.\ C {\bf 85}, 044911 (2012).

\bibitem{Chatterjee:2009ys} 
  R.~Chatterjee, D.K.~Srivastava and U.~Heinz,
  arXiv:0901.3270 [nucl-th].

\bibitem{Dion:2011pp} 
  M.~Dion {\it et al.}, 
  Phys.\ Rev.\ C {\bf 84}, 064901 (2011).

\bibitem{vanHees:2013}
H.~van Hees {\it et al.}, in preparation (2013).

\bibitem{Bzdak:2012fr} 
  A.~Bzdak and V.~Skokov,
  Phys.\ Rev.\ Lett.\  {\bf 110}, 192301 (2013).

\end{thebibliography}
\end{document}